# Virtual Data in CMS Production

A. Arbree, P. Avery, D. Bourilkov, R. Cavanaugh, S. Katageri, J. Rodriguez
*University of Florida, Gainesville, FL 32611, USA*

G. Graham
*FNAL, Batavia, IL 60510, USA*

J. Vöckler
*University of Chicago, Chicago, IL 60637, USA*

M. Wilde
*ANL, Argonne, IL 60439, USA*

Initial applications of the GriPhyN Chimera Virtual Data System have been performed within the context of CMS Production of Monte Carlo Simulated Data. The GriPhyN Chimera system consists of four primary components: 1) a Virtual Data Language, which is used to describe virtual data products, 2) a Virtual Data Catalog, which is used to store virtual data entries, 3) an Abstract Planner, which resolves all dependencies of a particular virtual data product and forms a location and existence independent plan, 4) a Concrete Planner, which maps an abstract, logical plan onto concrete, physical grid resources accounting for staging in/out files and publishing results to a replica location service. A CMS Workflow Planner, MCRunJob, is used to generate virtual data products using the Virtual Data Language. Subsequently, a prototype workflow manager, known as WorkRunner, is used to schedule the instantiation of virtual data products across a grid.

## 1. INTRODUCTION

The term "virtual data" as used in grid computing represents an abstraction of actual computational workflows, which are site specific and data existence dependent, to that of workflow descriptions, which are independent of both location and existence. Because the execution environment in grid computing is intrinsically both heterogeneous and dynamic, the use of virtual data provides flexibility to grid execution planners, allowing them to take advantage of scheduling techniques normally reserved for well understood, homogenous environments. From a physicist's point of view, however, virtual data represents a "recipe" for re-producing historical data or producing future data and, as a consequence, provides a provenance record of any data product created within a virtual data system.

This paper reports on the initial application of the GriPhyN Virtual Data System, known as *Chimera* [1], to the CMS Monte Carlo production of simulated data in a grid environment.

## 2. THE CHIMERA VIRTUAL DATA SYSTEM

The Chimera Virtual Data System requires that jobs be expressed in a Virtual Data Language (VDL). Chimera currently operates on two types of VDL objects: *transformations* and *derivations* [2]. A *transformation* is a schema defining formal types of input and output required to execute a particular application and is mapped onto an executable. A *derivation* represents a particular invocation of a transformation and is an actual virtual data product; it contains a detailed record of all the parameters needed to create a piece of actual data. Virtual data becomes physical data when its derivation is executed. A transformation thus represents a class of virtual data objects, while a derivation represents a particular virtual data object. [1, 3]

Chimera stores these VDL objects in a Virtual Data Catalogue (VDC). When a user requests that a particular virtual data product be instantiated, Chimera invokes an "abstract planner" which queries the VDC. This query recursively traverses the associated graph of derivations, thereby satisfying any required (virtual) data dependencies, and returns an abstract workflow plan in the form of a Directed Acyclic Graph (DAG), where each node represents an application and each edge an input/output dependency. This abstract DAG, the first step in the execution process, specifies all of the I/O dependencies required to create the requested virtual data product and represents the graph with maximal (virtual) data dependencies.

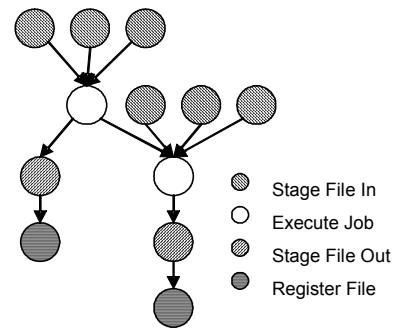

**Figure 1: A concrete Chimera DAG corresponding to a CMS virtual data request: Mapping an abstract DAG onto physical grid resources produces a concrete DAG, which can be directly submitted to the grid via Condor-G/DAGMan.**

Once an abstract workflow plan for the requested virtual data product has been produced, the plan is then mapped onto a set of actual grid resources for execution [4]. Given a user declared site for execution, Chimera invokes the Pegasus "concrete planner" [5] which first checks a Replication Location Service (RLS) [6] for the actual existence of any dependent input/output data in the abstract DAG, pruning any superfluous edges, and secondly dresses up any remaining nodes with stage-in, stage-out, and register nodes accounting for any location specific requirements of the chosen grid execution site (see Figure 1). The resulting concrete DAG then contains the minimal execution steps required to produce the requested virtual data product and is fully resolved onto particular grid resources. The Chimera Virtual Data System produces the concrete DAG in the Condor/DAGMan ClassAd format, enabling the user to directly submit the plan for the requested virtual data product to Condor-G/DAGMan [7, 8].

## 3. A CMS PRODUCTION USE-CASE OF SIMULATED DATA

A variety of different use-cases exist for production of simulated CMS data however, for this study, only one of the simpler cases was implemented: an "n-tuple-only" production, consisting of a five stage computational pipeline as shown in Figure 2.

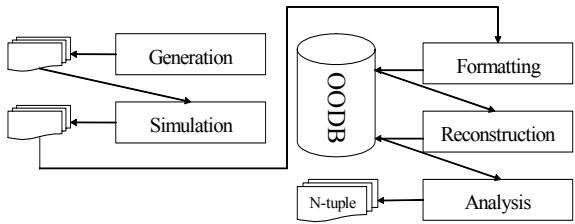

**Figure 2: N-tuple-only CMS simulation pipeline. The two stages on the left simulate the CMS detector and three stages on the right process events from the (simulated) detector.**

For simplicity, this particular pipeline can be split into two sections: simulation and reconstruction. The initial section consists of two FORTRAN application stages. The first, a generation stage called CMKIN, simulates the underlying physics of each event using Pythia [9]. The second stage, called CMSIM, corresponds to detector simulation and models the CMS experimental apparatus' response via GEANT 3 [10] to the events created in the first stage.

The second section, reconstruction and analysis[1], consists of three different application invocations of the CMS analysis and reconstruction C++ framework, known as ORCA [11]. Each ORCA application is a dynamically linked executable and requires local access to a CMS specific data persistency system.[2] The first ORCA application stage, hit formatting, copies the simulated detector raw hit data from a flat FORTRAN file into an object-oriented database. The next ORCA application stage reconstructs the physics event data from the raw detector response. The final ORCA application stage corresponds to analysis and selects (or possibly creates) user-specific information from the reconstructed event data, producing a smaller, concise, flat n-tuple file that is more easily analyzed by CMS physicists. This last file, in an n-tuple-only production, is the only important piece of data. All other intermediate files and databases may be discarded and only log files for the intermediate data products need to be kept for quality assurance validation.

## 4. APPLYING CHIMERA TO CMS PRODUCTION

For the chosen CMS production use-case, only one set of transformations, defining the input/output dependency schema for each application, had to be created. It would have been optimal to create one transformation for each stage in the CMS pipeline, but limitations of the CMS executables and the grid environment required that they be lumped into two transformations, one for each section of the pipeline (represented by the execution job nodes in Figure 1). The below text represents example VDL code for the CMKIN/CMSIM transformation and a corresponding derivation:

```
TR FORTRAN_SECTION( none runnum, none project,
        none numevents, output outfile,
        input kincard, input simcard,
        input geomfile, output logfile )
{
    argument = ${none:runnum};
    argument = ${none:project};
    argument = ${none:numevents};
    argument = ${input:kincard};
    argument = ${input:simcard};
    argument = ${input:geomfile};
    argument = ${output:logfile};
    argument = ${output:outfile};
}

DV EG02_BIGJETS_1_SIMULATION->FORTRAN_SECTION(
    kincard=@{input:"eg02_BigJets_Id_252.txt"},
    simcard=@{input:"STANDARD_125_Id_42.txt"},
    geomfile=@{input:"cms125.fz"},
    logfile=@{output:"fortran.eg02_BigJets_1.log"},
    numevents="250",
    outfile=@{output:"eg02_BigJets_1.fz"},
    project="eg02_BigJets",
    runnum="1" );
```

The transformation argument types `input`, `output`, `none` explicitly indicate dependencies. Chimera resolves input dependencies by RLS lookup or transformation invocation, and output dependencies by RLS register; arguments that pose no dependency requirements are simple parameters. In this particular example, `project` represents the name of the production; `runnum` corresponds to job-splitting (see below) and doubles as the random number seed used for the job, `numevents` represents the number of events which are to be produced by the job, `kincard` is the CMKIN input card file, `simcard` is the CMSIM input card file, `geomfile` specifies the input GEANT geometry file for the CMS detector, `logfile` names the output file for "stdout" from both the CMKIN

---

[1] Initial applications of the Chimera Virtual Data System with specific emphasis on the analysis of simulated CMS physics data have been performed and are reported in [14].

[2] At the time of this study, Objectivity/DB was used as the CMS data persistency system.

and CMSIM applications, and `outfile` identifies the output CMSIM data file.[3]

As Chimera requires that only one executable be mapped to a transformation, CMS application wrapper scripts were written to encapsulate the interaction of all the executables within a section. These scripts, in addition to encapsulating the various stages, must also gather information needed to update a centralised CMS reference metadata database (known as the RefDB [12]) and must configure the execution environment for the CMS executables. Configuring the local CMS execution environment was a significant overhead, requiring manual pre-installation and configuration of the CMS software at each individual grid site.

A typical CMS production run requires that hundreds, sometimes thousands, of derivations be created. In addition, CMS requires that all sites producing CMS simulations read their input parameters from the CMS metadata database (RefDB) and write back to the database a list of values describing how and when the data was produced [12]. In order to meet these requirements, several proof-of-concept Bourne Shell scripts were written.

First, prototype scripts were written to interface Chimera with the RefDB. There are two interface scripts: one reads from the database and one writes to the database. The "read" script queries the RefDB and produces a job description file. This file is used later to write the derivations. The "write" script parses the log files produced by the CMS wrapper scripts described above for information about the success of the job, the size and location of the output files, and many of the physical parameters created by the simulation; this parsed data is reformatted and updated into the RefDB.

The job description file produced by the "read" script is passed to a prototype derivation generating script. This script reads the job description file to determine various job parameters. A typical production may require 100,000 or more events to be produced, requiring several thousand CPU/hours to compute. Due to the parallel properties of the Monte Carlo simulation, this large production is split into a number of smaller jobs for computation. The "derivation generating" script performs this splitting and adjusts all of the input parameters accordingly, before writing out all of the derivations and creating the abstract DAG for each individual CMS virtual data product.

With the success of the initial study, the relevant functionality of the above RefDB interface scripts and the prototype derivation generating scripts have now been incorporated into the official CMS metadata and workflow manager, MCRunJob [13], which queries and updates the RefDB, performs the job-splitting and directly writes the appropriate Chimera derivations required for each virtual data product into the VDC.

## 5. A GRID SCHEDULER FOR VIRTUAL DATA PRODUCTION

In order to enable the Chimera system to support long running, production workflows on a virtual data grid [4], a prototype scheduler, known as WorkRunner was written (see Figure 3).

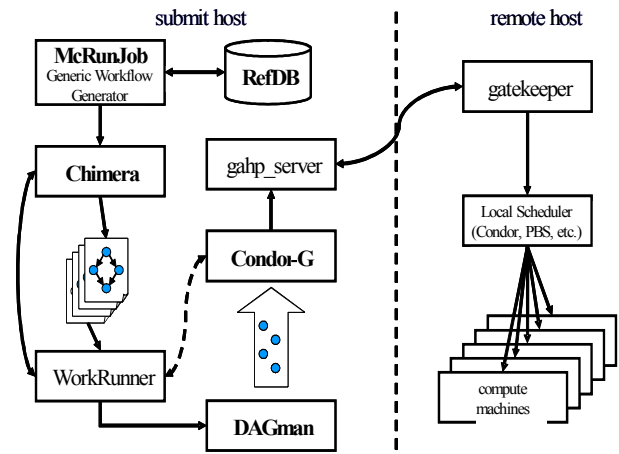

**Figure 3: Important pieces of the virtual data grid used by this study. The GriPhyN Virtual Data Toolkit was used to build the basic grid infrastructure.**

WorkRunner is invoked once all virtual data derivations have been pre-recorded into the VDC by the metadata and workflow manager (initially the prototype scripts described above and now by MCRunJob). Rather than immediately concretizing the DAGs and submitting the jobs to the grid, the abstract DAGs are queued into WorkRunner. WorkRunner determines which grid sites are free to receive work and submits DAGs from its queue to keep them working at maximum capacity. WorkRunner is divided into three modules. First, a Condor-G interface monitors information from the Condor-G submission program to determine which sites in the grid are in "need" of work. This "need" is determined using high and low watermarks. WorkRunner determines the number of jobs running at each site and if this number is below the low watermark, it submits jobs until it rises above the high watermark. The user supplies the high and low watermarks so they reflect the expected throughput of the particular grid site. Second, a Chimera interface takes queued abstract DAGs, asks Chimera to concretize them, and then submits them to the appropriate site through Condor-G. Finally, a job-tracking module records information about the status of each job in the WorkRunner queue as well as those, which have been sent and are executing via Condor-G.

---

[3] In principle one could store all of the CMKIN and CMSIM parameters contained in the input card files as VDL formal entries of type `none` in the transformation definition with actual values passed to every derivation. But, as the vast majority of these necessary parameters are static, they are currently stored as a dependent input file for reasons of economy. This argues for a rich, persistent mechanism for the manipulation of structured application metadata.

## 6. INITIAL RESULTS

This project had two testing phases, an integration test and a scheduling test. After the wrapper scripts were completed, all the VDL components were created, and the derivation generation scripts were written, a test production of 150,000 events was performed. This first test used the University of Florida GriPhyN computing cluster consisting of 25 dual-processor Pentium (1 GHz) machines. Over the course of 7 days, 678 DAGs each computing 250 events were submitted. Of these DAGs, 670 were successful and produced 167,500 events using approximately 350 CPU/days of computing power and producing approximately 200GB of simulated data. The 8 unsuccessful DAGs failed when accidentally preempted by local user.

The second test was designed to test the power of the new scheduler as well as the Chimera integration on a wider variety of computing sites. New sites were built at the University of Florida High Speed Simulation and Computing Center, the University of Chicago Computer Science Department, the University of Wisconsin, Milwakee, Physics Department, and at the Argonne National Laboratory DataGrid experiment cluster. Additionally, the resources of the University of Florida physics cluster were divided and a second cluster was built with half of the machines. All grid sites used the GriPhyN Virtual Data Toolkit as the base grid middleware. In addition, all required CMS software[4] and supporting scripts were pre-installed and configured on all grid sites.

For the scheduling test, a larger number of DAGs were used. In order to bring the computational time for each job down, only one event was produced with each DAG. However, 10,000 such DAGs were created. WorkRunner was used to submit these DAGs to all 6 of the compute sites. After continuously running for 4 days, 5954 DAGs were submitted to the grid, of which 5559 DAGs returned their data successfully. The 6.3% failure rate was higher than anticipated and resulted from unforeseen failures in the other grid components. In one instance, an entire compute site lost communication with the submitting site for several hours and approximately 200-300 hundred DAGs failed to write back their data back during this time. The cause for the remainder of the failed jobs, about one random job in fifty, appears to be due to known deficiencies in early versions of the grid middleware.

## 7. SUMMARY

Tests of the GriPhyN Chimera Virtual Data System have been performed within the context of CMS Production of Monte Carlo Simulated Data. A CMS Workflow Planner, MCRunJob, was used to generate virtual data products using the Virtual Data Language.

Subsequently, a workflow manager, known as WorkRunner, was used to schedule the instantiation of virtual data products across a grid. The tests demonstrate the feasibility of using such a Virtual Data System to form a production workflow using high level, existence and location independent descriptions as well as, automatically mapping and scheduling such high level, logical descriptions of workflow onto physical grid resources within an actual CMS production environment.

## Acknowledgments

We would especially like to thank our colleagues in the High-performance Computing & Simulation Research Lab at the University of Florida including Prof. Alan George, Raj Subramaniyn, and Sarp Oral for the use of their resources. In addition, the authors wish to particularly recognise Michael Milligan, Guarang Mehta and Karan Vahi, for many helpful discussions contributing to this work.

This work is supported in part by the United States National Science Foundation under grants NSF ITR-0086044 (GriPhyN) and NSF PHY-0122557 (iVDGL).

---

[4] As the ORCA application stages were only certified for execution within a Red Hat Linux 6 environment, only grid sites running that version of the operating system were used for running the ORCA stages of a DAG.